\documentclass[useAMS,usenatbib]{mn2e}

\usepackage{amsmath}
\usepackage{amssymb}
\usepackage{bm}
\usepackage{comment}
\usepackage{graphicx}
\usepackage{times}

\newcommand{\LSun}{\mbox{L}_\odot} 
\newcommand{\MSun}{\mbox{M}_\odot} 
\newcommand{\Mpyr}{\mbox{M}_\odot\,\mbox{yr}^{-1}} 
\newcommand{\pd}{\partial} 

\defcitealias{vorobyov2013}{VDB~2013}

\title[The Luminosity of Population III Star Clusters]{The Luminosity of Population III Star Clusters}
\author[A.~L.~DeSouza and S.~Basu]{Alexander L.~DeSouza and Shantanu
  Basu\thanks{e-mail: alexander.desouza@gmail.com, basu@uwo.ca} \\
Department of Physics and Astronomy, The University of Western
Ontario, 1151 Richmond Street, London, ON, Canada, N6A 3K7}

\begin{document}

\maketitle

\begin{abstract}
We analyze the time evolution of the luminosity of a cluster of
Population III protostars formed in the early universe. We argue from
the Jeans criterion that primordial gas can collapse to form a cluster
of first stars that evolve relatively independently of one another
(i.e., with negligible gravitational interaction). We model the
collapse of individual protostellar clumps using 2+1D nonaxisymmetric
numerical hydrodynamics simulations. Each collapse produces a
protostar surrounded by a massive disk (i.e., $M_{\rm disk} / M_{*}
\gtrsim 0.1$), whose evolution we follow for a further 30--40
kyr. Gravitational instabilities result in the fragmentation and the
formation of gravitationally bound clumps within the disk. The
accretion of these fragments by the host protostar produces accretion
and luminosity bursts on the order of $10^6\,\LSun$. Within the
cluster, we show that a simultaneity of such events across several
protostellar cluster members can elevate the cluster luminosity to
5--10${\times}$ greater than expected, and that the cluster spends
$\sim15\%$ of it’s star-forming history at these levels. This enhanced
luminosity effect is particularly enabled in clusters of modest size
with $\simeq$ 10--20 members. In one such instance, we identify a
confluence of burst events that raise the luminosity to nearly
$1000{\times}$ greater than the cluster mean luminosity, resulting in
$L > 10^8\,\LSun$. This phenomenon arises solely through the
gravitational-instability--driven episodic fragmentation and accretion
that characterizes this early stage of protostellar evolution.
\end{abstract}



\section{Introduction}
\label{sec:introduction}

Following the emission of the Cosmic Microwave Background, the
universe entered a cosmological ``dark age'' that ended with the
formation of the first stars in the universe. These first stars (known
as Population III stars) were responsible for producing the
ultraviolet radiation that began the reionization of the universe
\citep[e.g.,][]{tumlinson2000}, and their supernovae were responsible
for enriching the intergalactic medium with the first heavy elements
\citep[e.g.,][]{miralda-escude1997,gnedin1997,ferrara2000}.

Cosmological-scale simulations that follow both the dark matter and
baryonic components of the early universe have yielded the consensus
opinion that Population III stars formed in dark matter minihalos with
masses of approximately $10^6\,\MSun$. These $3\,\sigma+$
perturbations over the background dark matter density field virialized
by redshifts of $z$ $\sim$ 20--50
\citep{tegmark1997,abel2002,bromm2002}. With few exceptions
\citep[e.g.,][]{turk2009}, these simulations suggest that the gas
pooling into the halos underwent a quasi-hydrostatic contraction until
they had sufficient mass to trigger runaway gravitational collapse
\citep{abel2002,bromm2002,bromm2004a,yoshida2006,o'shea2007,yoshida2008}. These
studies established the standard paradigm that the progenitor cloud
cores of the first stars were most likely to have been massive and
formed in relative isolation.

In contrast, it is well understood (theoretically as well as
observationally) that most star formation in the present-day universe
arises from the fragmentation of molecular clouds, resulting in a
multiplicity of young stellar objects being formed in close proximity
to each other
\citep[e.g.,][]{carpenter1997,hillenbrand1997,lada2003}. Motivated by
this, several recent studies have explored the fragmentary nature of
primordial gas in the early universe, and have been able to resolve
fragmentation in the disk-like environments surrounding the first
protostars, thus challenging the standard paradigm \citep[][hereafter
VDB\,2013]{stacy2010,clark2011b,greif2011,vorobyov2013}.

Clearly some ambiguity remains regarding the initial conditions and
the formation mechanism(s) of the first stars. Observations will be
required to accurately distinguish between the many existing theories
of Population III star formation and evolution. In fact, the detection
of primordial star clusters and galaxies in the early universe has
already been defined as a major goal for next generation telescopes
\citep[e.g.,][]{windhorst2006}. \citet{bromm2001} were among the first
to investigate the spectral energy distribution of primordial stars
theoretically. Later studies have expanded on their results to show
that isolated Population III stars are likely to be too faint for detection
by instruments such as the forthcoming James Webb Space Telescope,
even when their fluxes are enhanced via chance gravitational lensing
\citep{rydberg2013}. Several authors have also turned their attention
toward the potential of observing clusters, dwarf galaxies, and
massive galaxies that contain Population III stars that may have
formed at lower redshifts (i.e., $z < 10$) due to inhomogeneous metal
enrichment of the intergalactic medium following the first supernovae
\citep[e.g.,][]{ciardi2001,scannapieco2003,tornatore2007,johnson2010,safranek-shrader2014}.

In this paper we propose a scenario for the formation of a cluster of
Population III stars. We argue that the gas pooling into the dark
matter halos in which the first stars formed is subject to the Jeans
criterion analogously to the fragmentation of giant molecular clouds
in the present-day universe. As a result, these halos were capable of
producing small clusters of first stars. Using nonaxisymmetric 
numerical hydrodynamics simulations, we study the
gravitational-instability--driven fragmentation and accretion in the
collapsing protostellar environment. The resulting burst mode of
accretion is even more prominent in a Population III environment than
in present-day star formation, as shown by
\citetalias{vorobyov2013}. We use the calculations for individual
cluster members to compile the frequency, magnitude, and luminosity of
burst events for each cluster as a whole. We find that a simultaneity
of accretion events can produce bursts of luminosity that are several
orders of magnitude greater than the mean cluster luminosity.

The structure of this paper is as follows. In Section 2 we describe
how gas that settles into the host dark matter halos is subject to the
Jeans fragmentation criterion, allowing for the formation of a cluster
of first stars. In Section 3 we describe our numerical simulations (as
well as our selections for the initial conditions) for the formation
of each protostar, and calculate the luminosity for each of these
cluster members. In Section 4 we discuss the implications of having a
multiplicity of protostars simultaneously experiencing bursts of
accretion, and calculate the effect this has on the luminosity of the
cluster. We also discuss the implications of this phenomenon for
future observational programs. Finally, in Section 5 we conclude with
a brief discussion of our results.


\section{The Case for Population III Star Clusters}
\label{sec:pop3starclustering}

The formation of a cluster of Population III stars is thought to begin
with the collapse of low density baryon gas into the gravitational
potential wells established by dark matter minihalos that have
collapsed and virialized by $z \sim$ 20--50. The gas to dark matter
fraction in these halos is roughly $10\%$, and amounts to a gas mass
of $10^4$ to $10^5\,\MSun$. Numerical simulations of this collapse
reveal that the gas streaming into these potential wells exhibits
filamentary and knotty structure
\citep[e.g.,][]{bromm1999,clark2008,greif2011}. Indeed, gravity is
well known to enhance such anisotropic structure during collapse
\citep[e.g.,][]{lin1965}.

The formation of $\mbox{H}_{\rm 2}$ cools the gas efficiently
(and lowering the Jeans mass) from temperatures of a few times
$1000\,\mbox{K}$ to a few times $100\,\mbox{K}$, allowing the gas
density to increase to $n \sim 10^4\,\mbox{cm}^{-3}$. While the
imprint of substructure exists within the gas morphology, the
formation of $\mbox{H}_{2}$ is inefficient for cooling the gas below a
temperature of a few times $100\,\mbox{K}$, inhibiting further
collapse. Instead, these precursor imprints of fragmentation must next
undergo a slow quasi-hydrostatic contraction as they accrete
additional mass
\citep[e.g.,][]{tegmark1997,abel2002,bromm2002}. Runaway gravitational
collapse is only then triggered when the mass of these ``weak clumps''
exceeds the local Jeans value at this scale, being
\citep[e.g.,][]{clarke2003} 
\begin{equation}
\label{eqn:pop3jeansmass}
M_{\rm J}
\simeq
400\,\MSun
\left( \frac{T}{300\,\mbox{K}} \right)^{3/2}
\left( \frac{n}{10^5\,\mbox{cm}^{-3}} \right)^{-1/2}.
\end{equation}
For example, a halo with a gas mass of $10^5\,\MSun$ and a $10\%$ star
formation efficiency would form a weak cluster with $\simeq 20$
members. Though this sequence of events differs slightly from that of
present-day star formation, the latter also envisions a multiplicity of
approximately Jeans mass fragments that form in close proximity to
produce a weak or strong cluster \citep[e.g., in Taurus and $\rho$
Ophiuchus;][]{onishi1998,johnstone2000}.

These resultant massive clumps, each containing roughly $400\,\MSun$
of gas, are the sites of first star formation within the halo. The
dynamical state of a typical clump formed in this way---its mass,
physical extent, temperature, and angular momentum---are all
determined by the collapse process. The initial conditions for the
further contraction of these clumps are therefore relatively well
constrained \citep[e.g.,][]{yoshida2003,yoshida2006}.

Authors such as \citet{abel1998} have estimated that the efficiency of
this fragmentation---with which the gas pooling into the dark matter
halo is assimilated into high-density clumps---could vary between as
little as a few percent to nearly $50\%$. As the larger clumps tend to
retain their individuality against dispersal and/or mergers, the final
evolutionary state of the halo gas is expected to be a small cluster
of isolated objects
\citep[e.g.,][]{clark2008,turk2009,greif2012}. However, owing to 
lingering ambiguities about the manner in which this occurs, we adopt
the definition of a ``cluster'' as being any association
of $2+$ stars forming from independent gaseous clumps that result from
the subfragmentation of primordially pristine gas that has pooled into
a single dark matter halo. Additionally, we concern ourselves with
only the formation period during which the cluster members have masses
below $40\,\MSun$. This allows us to assume that the predominant gas
fraction within the halo is not ionized by the star formation as the
cluster evolution proceeds. In fact, such conditions represent the
analogue of present-day so-called ``embedded'' clusters, in which more
than $80\%$ of the cluster members belong to the Class II/III
evolutionary phases, and the mass function of the cluster is assumed
to be no longer evolving \citep[e.g.,][]{gutermuth2009}. We also
assume that each clump is able to promptly form stars within
${\sim}1\,\mbox{Myr}$, or roughly the lifetime of the most massive
individual stars \citep[e.g.,][]{bond1984}.

Some caveats to our model assumptions are that we are assuming cluster
formation according to the standard hydrodynamic Jeans criterion, and
disk evolution in the hydrodynamic limit. Cluster formation may be
strongly affected by several other factors. Turbulence generated by
colliding flows may leave a strong initial imprint that dominates
fragmentation \citep{heitsch2006,heitsch2008}. Radiative feedback from
the very first massive stars can also affect the further fragmentation
of a cloud or disk \citep{stacy2012}. Furthermore, if a seed magnetic
field exists and is amplified by dynamo action during primoridal cloud
collapse \citep{turk2012}, then magnetic field effects can
significantly affect the Jeans scale \citep{ciolek2006,basu2009} and
affect global mass flow in the cloud \citep{price2008}. Magnetic
fields can also inhibit the formation of large disks or their
fragmentation in simulations of present-day star formation
\citep[e.g.,][]{mellon2008,commercon2010,machida2011,dapp2012,tomida2015},
and their effect in the Population III regime remains to be further
elucidated.


\section{Individual Cluster Members}
\label{sec:individualclustermembers}

The clumps that emerge from the fragmentation are assumed to be
relatively isolated. Competition for accretion between clumps (due to
protostellar crowding) and effects arising from gravitational
interactions are negligibly small 
\citep[e.g.,][]{turk2009,hirano2014}. We thus model the formation and
evolution of each cluster member with its own unique simulation---the
initial conditions of which stem from the arguments of the preceding
section. Here we provide details about the the numerical aspects of
our code, our specific choices of initial conditions, and investigate
the long-term behavior exhibited in a fiducially constructed model.

\subsection{Numerical Simulations}
\label{subsec:numericalsimulations}

We carry out numerical simulations of the gravitationally induced
collapse of the primordial gas in 2+1D, assuming a thin-disk
geometry. Our code is a modified version of that presented in
\citet{vorobyov2005b,vorobyov2006}. The hydrodynamic equations are
solved using a finite difference scheme with a time-explicit
operator-split solution, based on \citet*{stone1992}. A thorough
description of our code is presented in \citetalias{vorobyov2013}.

The mass and momentum transport equations in the thin disk limit can
be expressed as
\begin{gather}
\label{eqn:continuityeqn}
\frac{\pd{\Sigma}}{\pd{t}} +
\bm{\nabla} \cdot \left( \Sigma \, \bm{v} \right) = 0, \\
\label{eqn:momentumeqn}
\frac{\pd}{\pd{t}} \left( \Sigma \, \bm{v} \right) + 
\bm{\nabla} \cdot \left( \Sigma \, \bm{v} \otimes \bm{v}  \right)
= - \bm{\nabla} P + \Sigma \, \bm{g},
\end{gather}
in which $\Sigma$ is the surface mass density, $\bm{v}$ is the
velocity of the disk material, $P$ is the vertically integrated
pressure (determined assuming the disk is in vertical hydrostatic
equilibrium at all times), and $\bm{\nabla}$ is the planar gradient
operator. The gravitational acceleration in the plane of the disk
($\bm{g}$) includes the contribution from the protostar (once
formed), from material inside the sink cell, and from the self-gravity
of the disk and surrounding cloud core. All vectoral terms and
quantities are understood as having only $\bm{\hat{r}}$ and
$\bm{\hat{\phi}}$ components in this formulation.

In order that the environment of the collapsing core accurately
reflect primordial conditions, equations (\ref{eqn:continuityeqn}) and
(\ref{eqn:momentumeqn}) are closed with a barotropic relation that
fits the 1D core collapse simulations of \citet{omukai2005}. These
include the detailed chemical and thermal processes of the collapsing
gas. Additional details of our model are provided in
\citetalias{vorobyov2013}.

\subsection{Initial Conditions}
\label{subsec:initialconditions}

The initial conditions for the radial gas surface density $\Sigma$ and
angular velocity $\Omega$ profiles for primordial cores are taken to
be very similar to those of present-day star forming cores
\cite[e.g.,][]{omukai1998,vorobyov2006,yoshida2008,vorobyov2010},
\begin{gather}
\label{eqn:initialsigmaprofile}
\Sigma = \Sigma_0 \left( 1 + \left( \frac{r}{r_0} \right)^2 \right)^{-1/2}, \\
\label{eqn:initialomegaprofile}
\Omega =
2\Omega_0 \left( \frac{r_0}{r} \right)^2
\left( \sqrt{ 1 + \left( \frac{r}{r_0} \right)^2 } - 1 \right).
\end{gather}
The radial surface mass density profile $\Sigma$ is that of an
integrated Bonnor-Ebert sphere \citep{dapp2009}, while the form of the
initial angular velocity profile $\Omega$ corresponds to the
differential rotation profile expected for a core collapsing out of a
near-uniform initial surface density field \citep{basu1997}.

We constrain $\Sigma_0$ by assuming a constant ratio of the radius of
the outer computational boundary $r_{\rm out}$ to that of the
centrally plateaued region $r_{\rm 0}$: $r_{\rm out} / r_0 \equiv 6$;
so that each core has a similar initial form with $\Sigma_0 \approx
0.25\,\mbox{g\,cm}^{-2}$. The parameter $r_{\rm out}$ thus also
determines the mass of the core, which for $r_{\rm out} =
0.5\,\mbox{pc}$ is approximately $300\,\MSun$. These choices are
constant for each of the cores simulated, and is consistent with the
typical size and mass found from ab initio cosmological simulations of
the collapse of primordial starless cores
\citep[e.g.,][]{yoshida2006}.

The angular momentum of the cloud core is parameterized by
$\Omega_0$---as appears in equation
(\ref{eqn:initialomegaprofile})---and is related to the dimensionless
parameter $\eta = ( \Omega_0 r_0 / c_{\rm s} )^2$
\citep{basu1997}. $\eta$ is related to the ratio of the rotational to
gravitational energy of the cloud core, $\beta = E_{\rm rot} /
|E_{\rm g}|$, with $\beta = 0.9 \eta$. Finally, $\beta$ relates to the
so-called ``spin parameter,'' $\lambda = \sqrt{\beta}$, that is most
often used to characterize the angular momentum of dark matter halos
(and their associated gas) formed from cosmological initial conditions
\citep[e.g.,][]{barnes1987,ryden1988}. The spin parameters of each
cloud core are lognormally distributed with mean $\bar{\lambda} =
0.05$ and variance $\sigma_{\rm \lambda}^2 = 0.5$
\citep[following][]{gardner2001,o'shea2007}.

\begin{figure*}
\centering
\includegraphics[width=\textwidth]{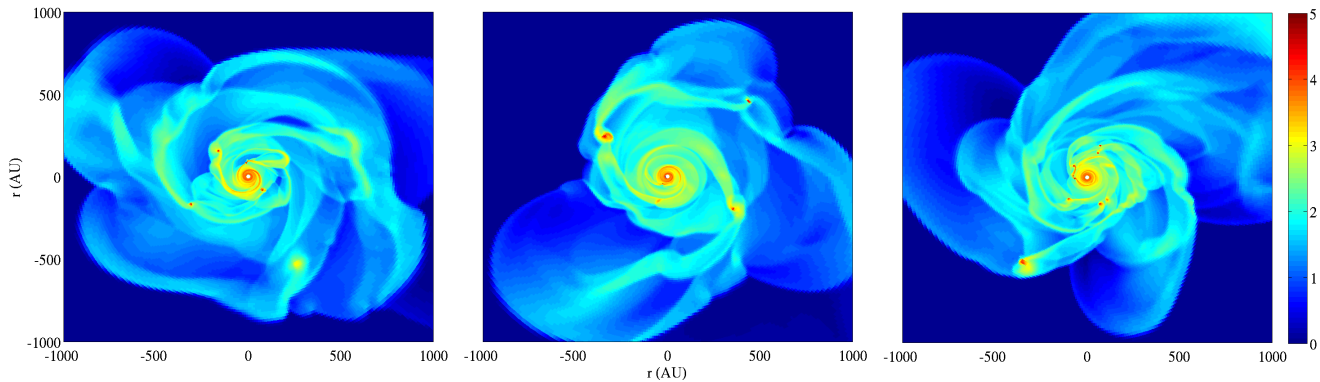}
\caption{Projection of the disk surface mass density $\Sigma$
  within a $2000 \times 2000\,\mbox{AU}$ volume centered on the
  accreting protostar. The time in each frame is $t = 5$, $10$, and
  $15\,\mbox{kyr}$ after formation of the protostar. \textbf{Left:}
  Some regions of the disk are
  Toomre-unstable, and several fragments can be identified at radii
  throughout the disk, between 10 and several hundred
  AU. \textbf{Center:} As these clumps are accreted onto the protostar
  or dispersed, the disk enters into a moderately quiescent period
  during which it is relatively stable against fragmentation. No
  fragments are present in the inner $300\,\mbox{AU}$ of the disk. The
  three large fragments at radii beyond this are remnants from an
  earlier phase of fragmentation that had been raised to higher
  orbits. \textbf{Right:} Continued accretion from the parent cloud
  core builds up the mass of the disk, eventually triggering another
  period of vigorous fragmentation. The two fragments immediately to
  the left of the sink cell in this panel provide an example of how
  larger fragments can be sheared apart prior to being accreted
  through the sink cell, leading to intense and rapid variability in
  the accretion luminosity of the protostar.}
\label{fig:threepanelplot}
\end{figure*}

Each model is run on a polar coordinate grid with $512 \times 512$
spatial grid zones in $r$ and $\phi$. The inner and outer boundary
conditions allow for free outflow from the computational
domain. Radial grid points are logarithmically distributed to allow
for better numerical resolution toward the innermost region of the
disk: the innermost cell outside of the sink region has a radius of
approximately $0.1\,\mbox{AU}$ and is $1.9\,\mbox{AU}$ (both radially
and azimuthally) at a radius of $100\,\mbox{AU}$.

\citet{tan2004} and \citet{hosokawa2011} have shown that beyond 30--40
$\MSun$ the role of increasing stellar luminosity becomes
critically important to understanding the subsequent evolution of
these protostars, as the intensely ionizing radiation begins to
inhibit $\mbox{H}_{\rm 2}$ formation, which is the primary coolant in
the gas. We therefore terminate our simulations once the protostar
reaches a mass of $40\,\MSun$.

\subsection{Evolution of the Protostellar Disk}
\label{subsec:evolutionoftheprotostellardisk}

Our fiducial model is characterized as a clump of gas roughly
$0.5\,\mbox{pc}$ in radius, with a mass of ${\sim}300\,\MSun$, and at
a temperature of $300\,\mbox{K}$. The spin parameter $\lambda$ of the
clump is equal to the mean of the distribution, $\bar{\lambda} \simeq
0.05$. The collapsing cores we model compare well to those derived
from 3D numerical simulations inside primordial minihalos
\citep[][]{yoshida2006,clark2011b}. Differences are attributable to
our cores being toward the lower end of the mass spectrum that has
been studied by these authors.

A quasi-Keplerian disk forms around the protostar within
${\sim}3\,\mbox{kyr}$ of the formation of the central protostar (in
our fiducial model). The disk begins to fragment within a few hundred
years after its formation. This timescale is somewhat longer than that
found by \citet{clark2011b} and \citet{greif2011}, who used sink cells
with smaller radii of $1.5\,\mbox{AU}$. However, this timescale is
similar to that found by \citet{smith2011}, who used sink cells with
comparable radii of $20\,\mbox{AU}$ (the central sink cell in our
simulation being ${\sim}10\,\mbox{AU}$ in radius).

In Figure \ref{fig:threepanelplot} we present three snapshots of the
disk surface density inside of a $1000\,\mbox{AU}$ radius and spanning
$10\,\mbox{kyr}$ of the disk evolution. The left panel shows that a
rich density structure already exists within the disk a mere 5 kyr
after formation of the central protostar. Together with the right
panel, the clearly defined condensations of gas that have fragmented
out of the disk indicate that the disk experiences multiple episodes
of instability. The central panel presents an intermediary quiescent
phase during which there is little to no fragmentation and the
accretion of gas onto the protostar occurs relatively smoothly. The
number of fragments within the disk clearly varies with time, as some
fragments are tidally dispersed and others are accreted onto the
central protostar; a result of the gravitational torques exerted by
the fragments on each other and from larger spiral arm structures that
form within the disk. Though accretion gradually drains the disk, new
episodes of fragmentation are continually stimulated by the resupply
of fresh gaseous material from the surrounding core envelope. It is
also evident that most of the fragments are formed in the intermediate
to outer disk region ($\gtrsim 50\,\mbox{AU}$), which is consistent
with the numerical simulations of disks around protostars in the
present-day universe \citep[e.g.,][]{stamatellos2008,clarke2009}.

\begin{figure}
\centering
\includegraphics[width=\columnwidth]{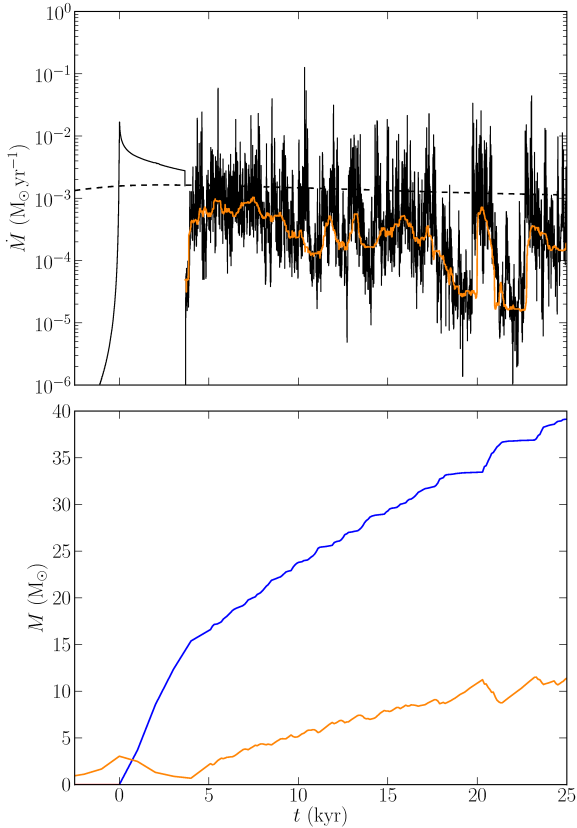}
\caption{\textbf{Top:} Temporal evolution of mass accretion rates:
  from the cloud core onto the disk at $3000\,\mbox{AU}$ (dashed
  black), and from the disk onto the protostar (solid black). The
  rectangular window time-averaged (in 1 kyr intervals) mass accretion
  rate---the quiescent accretion rate---is in orange. \textbf{Bottom:}
  Growth of the protostellar (blue) and disk (orange) masses in
  time. The protostellar mass grows via punctuated equilibrium, while
  the episodic increases in the mass of the protostar are coincident
  with the episodes of decline in disk mass. Nevertheless, the disk
  mass increases with time owing to the continually replenishment of
  disk material from the surround core material.}
\label{fig:accretion_fiducialmodel}
\end{figure}

Those fragments that pass through the inner computational boundary are
assumed to be readily accreted by the central protostar. Though this
boundary is not the protostellar surface itself, we can use the
temperature (or more appropriately, the sound speed) of the disk
material along this boundary to estimate the instantaneous mass
accretion rate onto the protostar as 
\begin{equation}
\dot{M} \approx \frac{ c_{\rm s}^3 } { G }
\end{equation}
\citep[][]{shu1977}.
In our fiducial model we find $c_{\rm s}$ is about
$2.0\,\mbox{km\,s}^{-1}$. This corresponds to an instantaneous mass
accretion rate of ${\sim}10^{-3}\,\Mpyr$. We note that this is just
below the critical (Eddington) accretion rate of $\dot{M}_{\rm  Edd}
\simeq 4.0 \times 10^{-3}\,\Mpyr$ above which the corresponding
radiation pressure alone is capable of halting the accretion flow
entirely \citep{hosokawa2009}. This estimate matches the time averaged
value of the mass accretion rate experienced by our model protostar
(Figure \ref{fig:accretion_fiducialmodel}), and is consistent with
several other models of the runaway collapse of primordial gas as has
been determined both analytically and from simulations
\citep[][]{omukai2003,bromm2004a,clark2011b,hosokawa2011,smith2012}.

We present the complete mass accretion history for our fiducial model
in the top panel of Figure \ref{fig:accretion_fiducialmodel}. The
formation of the protostar is marked by the sharp rise in the mass
accretion rate, which we define as time $t = 0$. Following this, material
from the surrounding envelope of the progenitor cloud core continues
to stream onto the protostar at a rate of a few times
$10^{-3}\,\Mpyr$. The mass of the protostar rapidly increases (within
${\sim}4\,\mbox{kyr}$) to approximately $15\,\MSun$ before the
accretion flow is temporarily halted by the formation of a
quasi-Keplerian disk.

The disk modulates the subsequent accretion of disk material onto the
protostar, with gravitational torques redistributing the mass and
angular momentum of the infalling cloud core material. In
\citetalias{vorobyov2013} we demonstrated that the disk self-gravity
quickly induces the formation of spiral arms. Furthermore, sections of
the disk in which the local value of Toomre's $Q$ falls below unity
become subject to fragmentation \citep{toomre1964}. These fragments
are then torqued inward along ballistic trajectories before ultimately
being accreted by the protostar, resulting in the episodic bursts of
accretion (see Figure \ref{fig:accretion_fiducialmodel}).

The cumulative effect of a quiescent mode of accretion that is
punctuated by the episodic bursts is also evident in the curves of
mass growth in the bottom panel of Figure
\ref{fig:accretion_fiducialmodel}. The protostellar mass (shown in
blue) grows rapidly during the initial phase of smooth accretion
(which lasts ${\sim}4\,\mbox{kyr}$). However, the burst mode of
accretion comes to dominate the subsequent growth of the protostar as
is evident by the abrupt increases in the protostellar mass, typically
a few $\MSun$ at a time. These increases are typically followed by
plateaus during which the mass of the protostar changes very little as
the disk equilibrates back into a quasi-Keplerian state. Each burst
event is mirrored by a corresponding decrease in the total disk mass
(in orange). However, the overall mass of the disk actually continues
to increase in time due to the steady accretion of material from the
remnant of the progenitor cloud core (the dashed black line in the top
panel of Figure \ref{fig:accretion_fiducialmodel}).

Additional details of the effects of the gravitational torques on the
disk, and the resulting recurrent character of the disk fragmentation,
are discussed in \citetalias{vorobyov2013}. Here we focus on the
signature of this behavior on the protostellar accretion luminosity.

\subsection{Accretion Luminosity}
\label{subsec:accretionluminosity}

A protostar's luminosity is a product of competition between mass
growth from accretion and radiative loss from the protostellar
interior. However, it is not until the protostar begins contracting
toward the main sequence that its internally generated luminosity
$L_*$ surpasses the accretion luminosity $L_{\rm acc}$. During the
earliest stages of its evolution, the source of a protostar's
luminosity is almost entirely from accretion, so that $L \approx
L_{\rm acc}$ \citep[up to masses of $M_* \simeq$ 30--40
$\MSun$;][]{tan2004,hosokawa2011}. In \citetalias{vorobyov2013} we 
showed that the large variability in accretion experienced by the
protostar results in accretion luminosities several orders of
magnitude greater than might otherwise be expected (compare
\citealt{clark2011b}, \citealt{hosokawa2011}, and \citealt{smith2012}
to \citetalias{vorobyov2013}, for example).

We estimate the accretion luminosity assuming that any material
landing on the surface of the protostar has its kinetic energy
dissipated radiatively at a rate
\begin{equation}
\label{eqn:Lacc}
L = \frac{ G M_* \dot{M} }{ 2 R_* },
\end{equation}
where $R_*$ is the protostellar radius.

In the absence of a detailed model for the stellar interior we instead
fit the evolutionary models of \citet{omukai2003} with a piecewise
power-law approximation to the radial expansion of the protostellar
surface as a function of the protostar mass. Following formation of
the hydrostatic core, the protostellar radius is expected to grow
according to a mixed power-law as $R_* \, {\propto} \, M_*^{0.27}
\,\dot{M}_{-3}^{0.41}$; where $\dot{M}_{-3}$ denotes the
ratio of the actual instantaneous mass accretion rate $\dot{M}$ to a
value of $10^{-3}\,\Mpyr$ \citep{stahler1986,omukai2003}. Increasing
temperature within the core drives a luminosity wave outward, causing
a rapid expansion of the stellar surface. When this wave breaches the
protostellar surface, the interior is able to relax and the protostar
begins Kelvin-Helmholtz contraction toward the main-sequence.

The following relations approximate the evolution of $R_*$ through
its transitions through these phases \citep{smith2011}:
\begin{equation}
\label{eqn:R-relations}
R_* = \left\{
\begin{array}{ll}
26 \, M_*^{0.27} \, \dot{M}_{-3}^{0.41}, & M_* \leq M_1 \\
A_1 \, M_*^3, & M_1 < \, M_* < M_2 \\
A_2 \, M_*^{-2}, & M_* \geq M_2 \, \mbox{and} \, R_* < R_{\rm MS}
\end{array} \right..
\end{equation}
The constants $A_1$ and $A_2$ are matching conditions that ensure the
functional form of $R_*$ is smoothly varying between transitions. The
mass parameter $M_1$ marks the transition between the adiabatic phase
of growth and the arrival of the luminosity wave at the protostellar
surface; $M_2$, the transition between the luminosity wave driven
expansion and subsequent Kelvin-Helmholtz contraction. $M_1$ and $M_2$
are fixed by the instantaneous mass accretion rate as the protostar
transitions between phases, and are defined as
\begin{equation}
\begin{array}{c}
M_1 = 5 \, \dot{M}_{-3}^{0.27}, \\
M_2 = 7 \, \dot{M}_{-3}^{0.27}.
\end{array}
\end{equation}
Note that as the evolution of the protostellar interior occurs roughly
adiabatically, due to the long cooling time therein, the variability
in the accretion rate induced by the burst mode of accretion does not
result in significant variability in the radius of the protostar.

\begin{figure}
\centering
\includegraphics[width=\columnwidth]{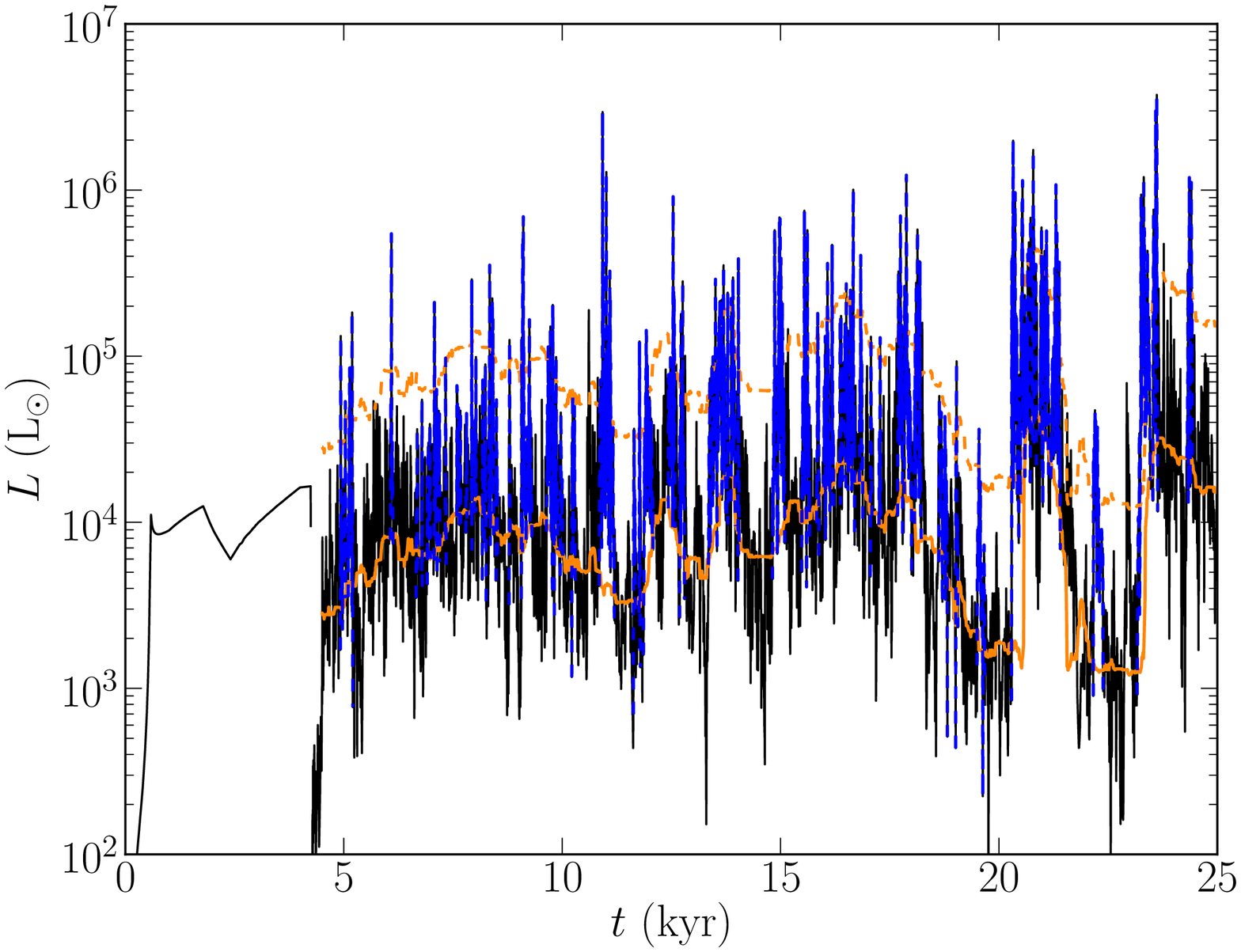}
\includegraphics[width=\columnwidth]{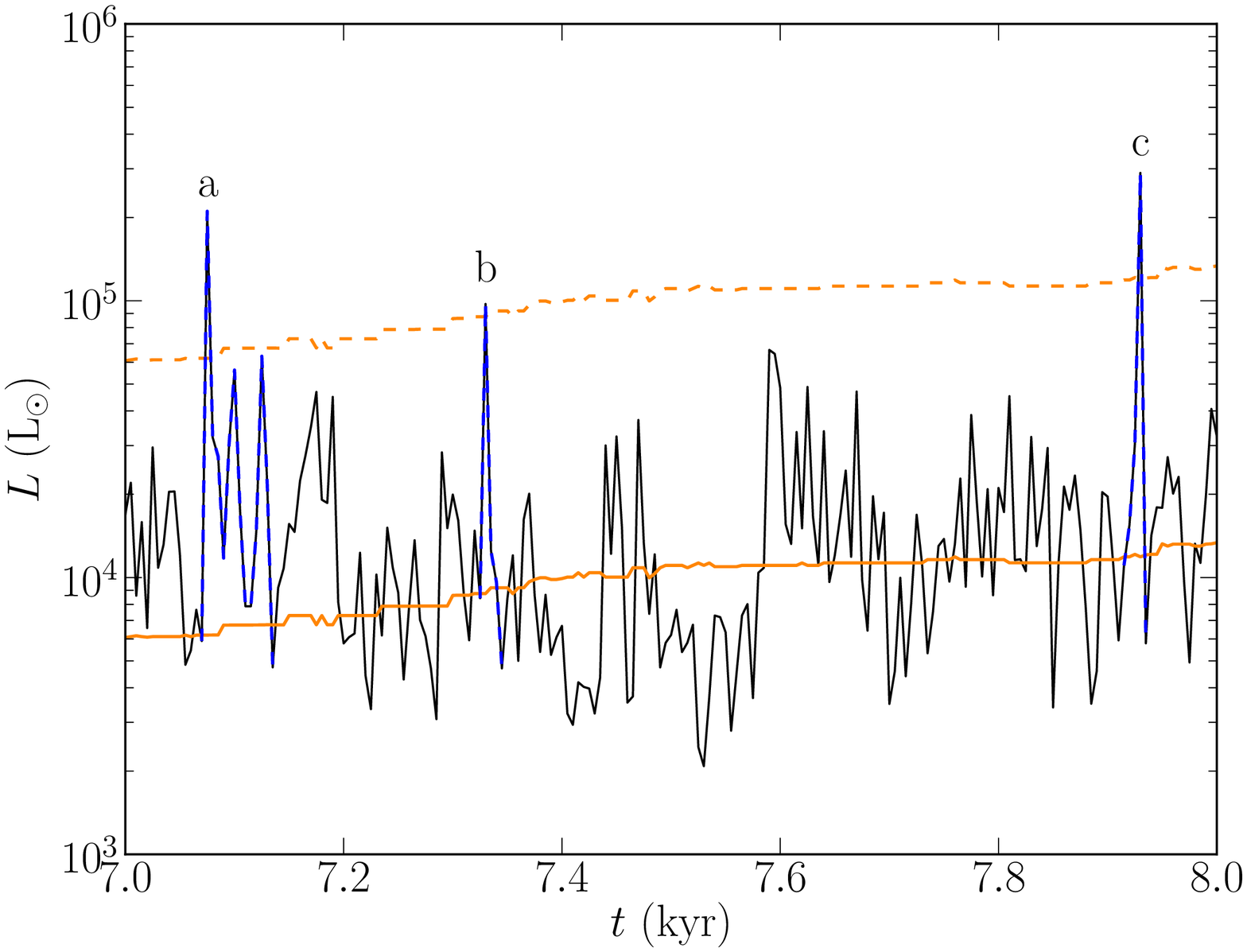}
\caption{\textbf{Top:} Temporal evolution of the (accretion) luminosity
  from the protostar in our fiducial model (in black). In overlay is
  the time averaged quiescent rate (solid orange), and the
  $10{\times}L_{\rm q}$ rate (dashed orange), with those accretion
  events whose peak luminosity exceeds this threshold highlighted in
  blue. \textbf{Bottom:} A zoom in on a $1\,\mbox{kyr}$ window of the
  luminosity history for clarity. Accretion events in excess of
  $10{\times}L_{\rm q}$ are clearly identified (labeled, a, b, and c,
  and highlighted in blue).}
\label{fig:Lacc_fiducialmodel}
\end{figure}

The accretion luminosity calculated for our fiducial model is shown in
Figure \ref{fig:Lacc_fiducialmodel} (black). In overlay is the time
averaged rate (i.e., the quiescent rate, $L_{\rm q}$, in orange), and a
demarcation $10{\times}$ greater than the quiescent rate (dashed
orange). Accretion events during which $L$ exceeds $L_{\rm q}$ by the
factor of $10$ are highlighted (dashed blue). Once the protostar is
formed (designated as $t = 0$), $L$ climbs very quickly to
${\sim}10^{4}\,\LSun$. The luminosity remains at about this level
during the period of smooth accretion until a disk is formed at $t
\sim 4\,\mbox{kyr}$. Although the mean rate remains at roughly
$10^4\,\LSun$, the episodic nature of the subsequent vigorous
accretion gives rise to significant variability, with some peaks
reaching several times $10^6\,\LSun$---two orders of magnitude greater
than $L_{\rm q}$. One might expect this large variability in
luminosity caused by the individual burst events to affect the
accretion flow via feedback from the enhanced radiation
field. However, as the duration of the individual bursts are quite
short (typically on the order of a few times $100\,\mbox{yr}$), we
expect that they do not have any appreciable effect on the long-term
growth of the protostar.

In the bottom panel of Figure \ref{fig:Lacc_fiducialmodel} we focus on
a $1\,\mbox{kyr}$ window of our simulation in order to highlight our
burst identification scheme. We calculate the effective mean accretion
rate (the solid orange line) using a moving $1\,\mbox{kyr}$ window. As
burst events are ultimately attributable to the accretion of large
individual fragments (with masses on the order of 0.1--1 $\MSun$) and
have an overall mean duration of a few times 100 yr, the use of a 1
kyr moving window allows us to supress variations due to the large
fragments being sheared apart into several smaller ones, each of which
is accreted by the protostar in rapid succession
\citepalias{vorobyov2013}. Hence, we also include as part of a single
burst event all points left- and rightward of the peak luminosity that
are greater than the mean (and not only those points for which $L$ is
strictly $> 10{\times}L_{\rm q}$).


\section{Clusters of Population III Protostars}
\label{sec:luminosityofpop3protoclusters}

\begin{figure*}
\centering
\includegraphics[width=0.485\textwidth]{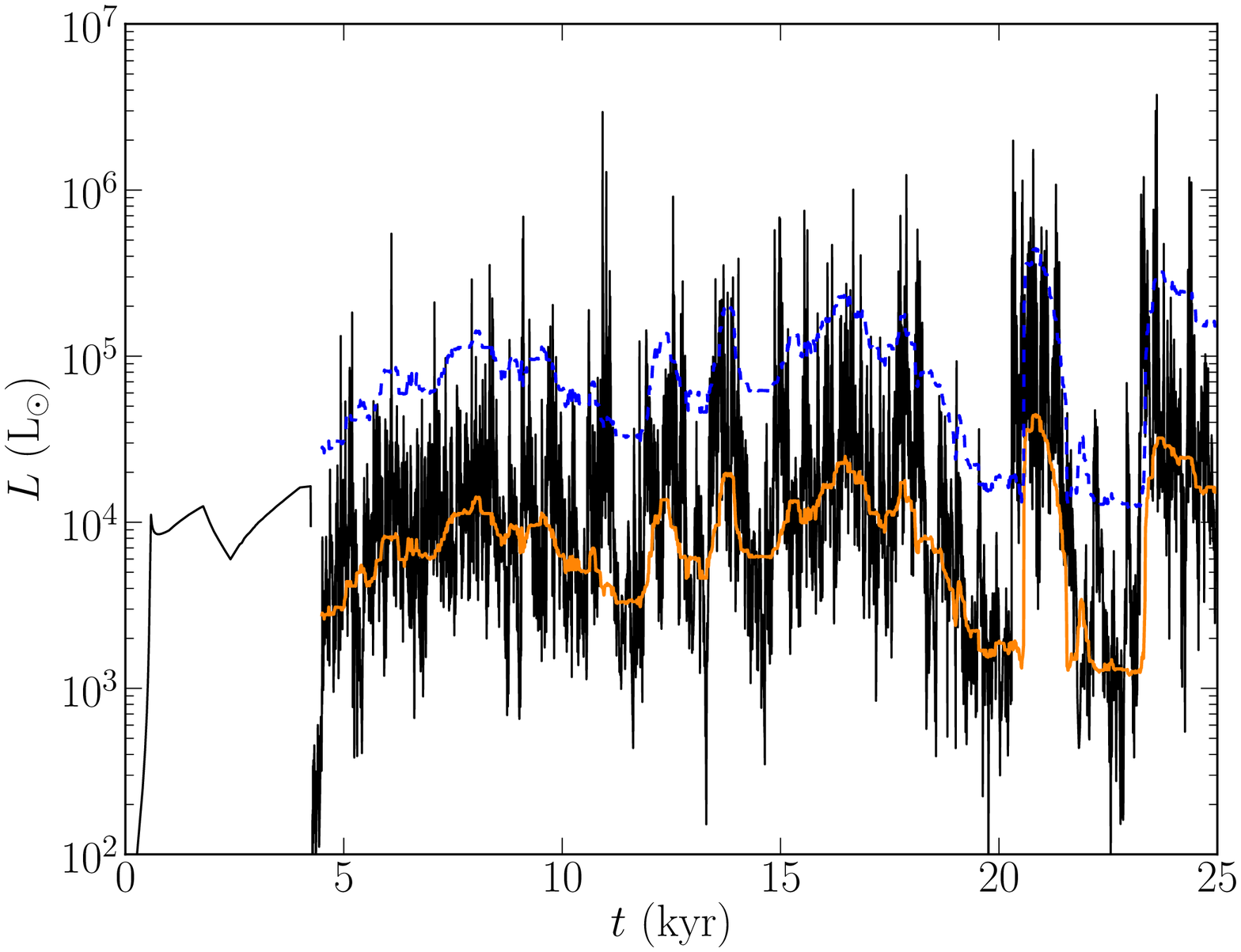}
\includegraphics[width=0.485\textwidth]{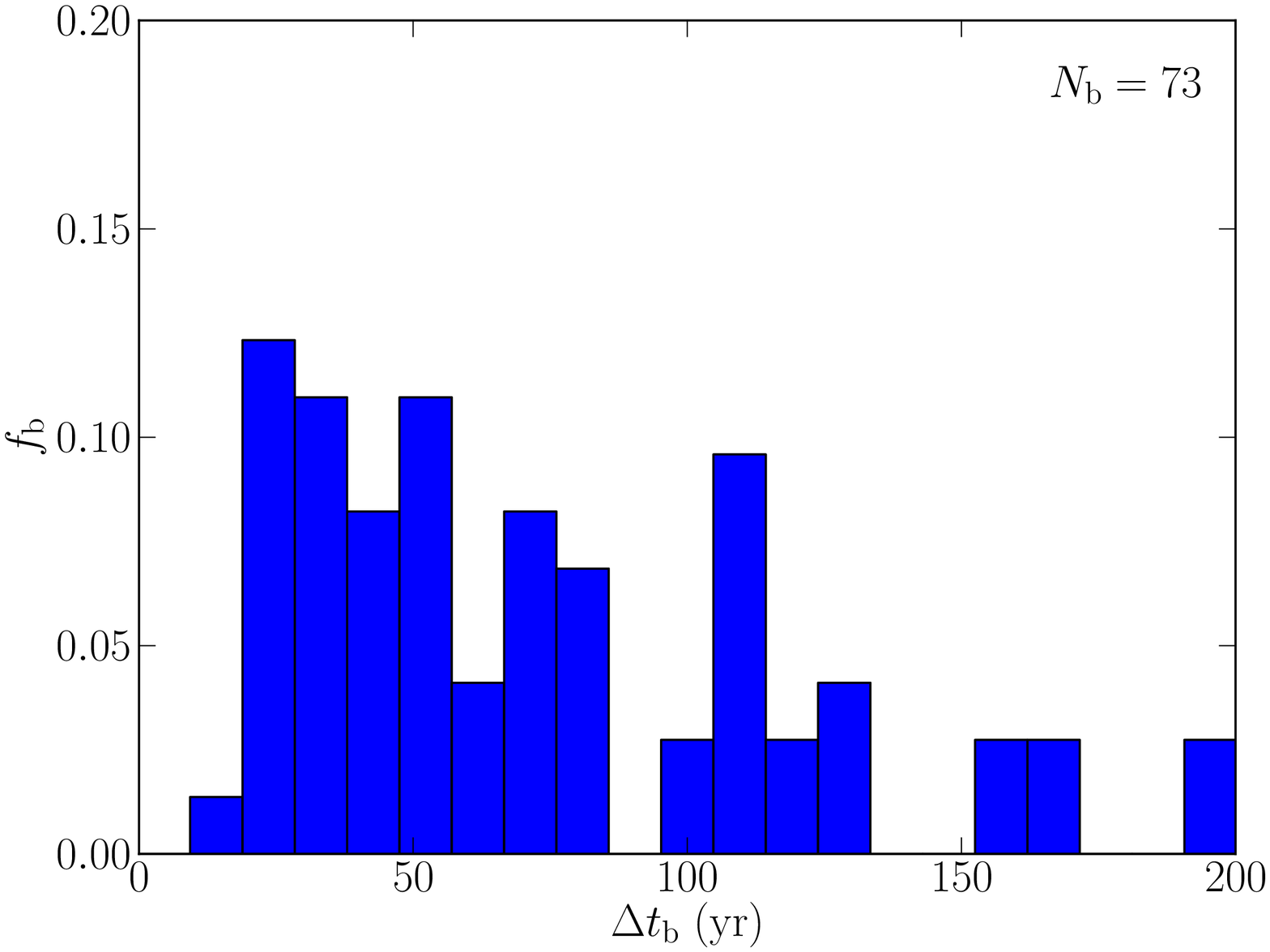} \\
\includegraphics[width=0.485\textwidth]{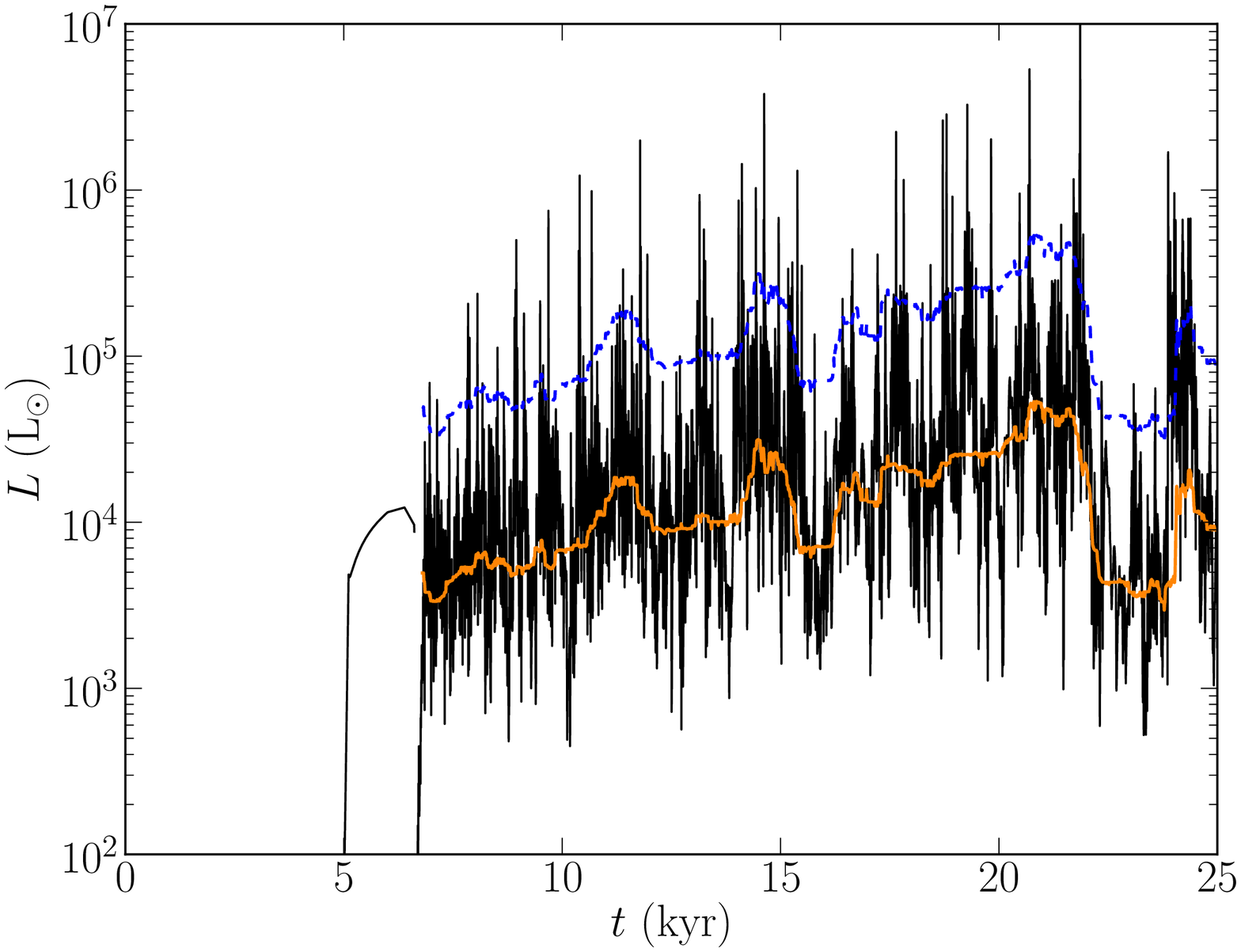}
\includegraphics[width=0.485\textwidth]{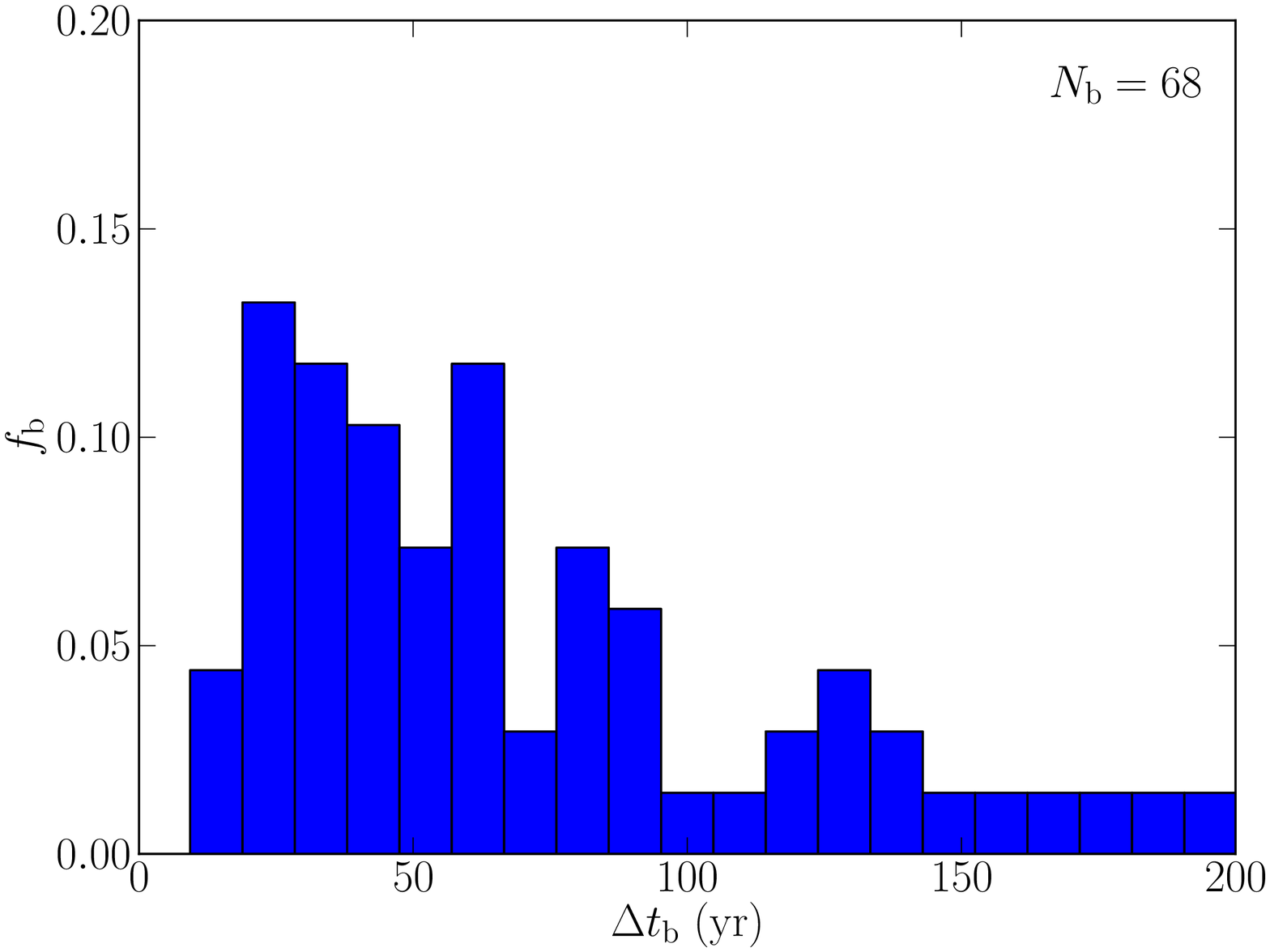} \\
\includegraphics[width=0.485\textwidth]{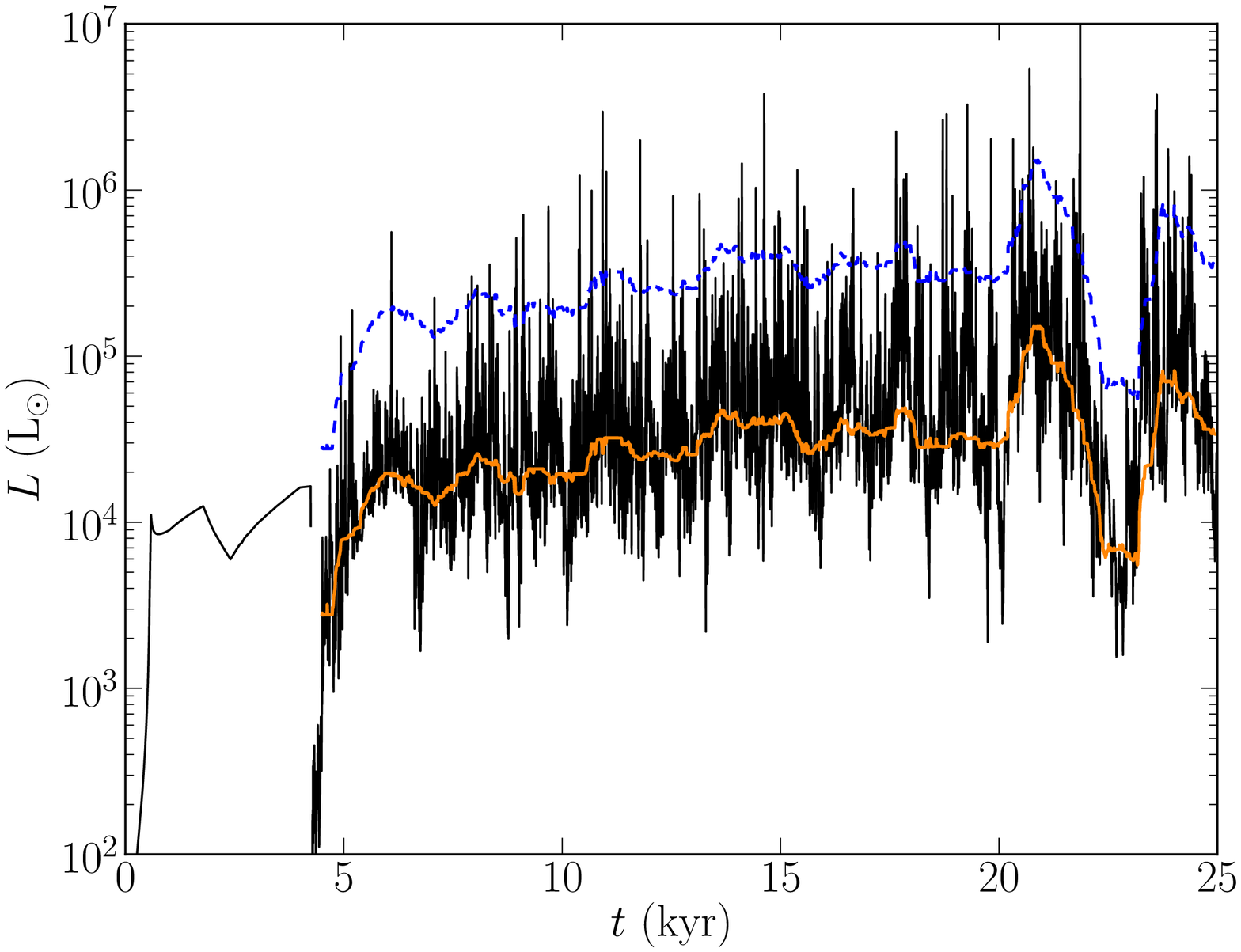}
\includegraphics[width=0.485\textwidth]{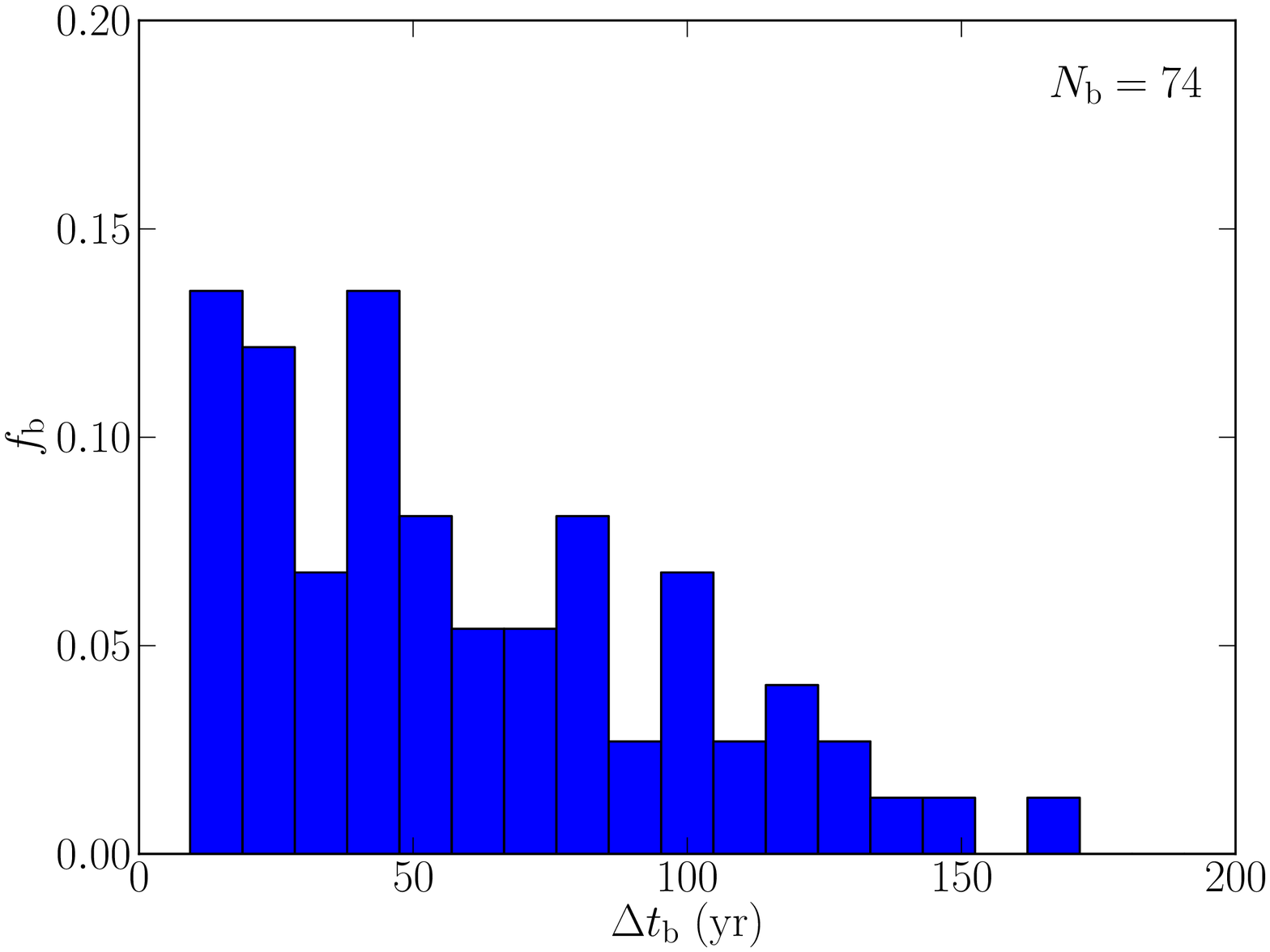}
\caption{Accretion luminosity $L$ and distribution of burst durations
  of the individual cluster members (top and middle panels) and for
  their combined output (bottom panel) in our example $N = 2$ member
  cluster. The top panel is our fiducial model with $\lambda = 0.050$;
  the middle panel is a model with $\lambda = 0.065$. \textbf{Left:}
  Accretion luminosities $L$ are in black; the quiescent luminosity
  $L_{\rm q}$ is in orange; the dashed blue line indicates a
  luminosity $10{\times}L_{\rm q}$. Note that the top panel is the
  fiducial model also shown in Figure
  \ref{fig:Lacc_fiducialmodel}. \textbf{Right:} Fractional number
  distributions ($f_{\rm b}$) of burst durations ($\Delta{t}_{\rm b}$)
  for the individual cluster members (top and middle), and for the
  cumulative cluster profile (bottom). The total number of identified
  bursts ($N_{\rm b}$) is indicated in the upper-right of each
  panel. The approximate mean burst duration observed for the
  individual cluster members are $96.3$ and $80.2\,\mbox{yr}$,
  respectively. The mean burst duration for the cumulative cluster
  luminosity profile is ${\sim}72.0\,\mbox{yr}$.}
\label{fig:N=2clusterexample}
\end{figure*}

Figure \ref{fig:N=2clusterexample} illustrates how the burst mode in
individual protostars can sum to produce a cluster luminosity. For
simplicity, we consider a cluster of only two stars, with one being
our fiducial model with $\lambda = 0.05$, and the second characterized
by $\lambda = 0.065$ but forming $5\,\mbox{kyr}$ after the first. The
panels in the top- and middle-left column are the luminosities of the
individual cluster members while the bottom-left shows the cumulative
luminosity for the cluster as a whole. The histograms on the
right-hand side present statistics of the fractional number
distribution of burst events $f_{\rm b}$ of different duration $\Delta
t_{\rm b}$. These can be used as metrics to evaluate differences
between the individual and cluster luminosities.

Prior to $t = 5\,\mbox{kyr}$, before the second cluster member forms, 
the luminosity of the cluster is of course just the luminosity of the 
first and as yet sole member. The second cluster member begins its
evolution from $t \sim 5\,\mbox{kyr}$ (where time $t = 0$ marks the
formation of the first cluster member). In the subsequent
${\sim}2\,\mbox{kyr}$, as the second cluster member smoothly accretes
material from its surroundings, its contribution to the total cluster
luminosity increases. In fact, comparing the cluster luminosity to the
individual component luminosities reveals that a significant amount of
the variability in the luminosity of the first cluster member is being
completely obscured. During this time, the cluster luminosity is
consistently above ${\sim}10^4\,\LSun$. By $t \sim 7\,\mbox{kyr}$,
both cluster members harbor their own protostellar disks, and
accretion onto each protostar is being driven by the action of
gravitational torques in the massive disks surrounding each
host---i.e., via the burst mode of accretion. Correspondingly, the
cluster luminosity thereafter exhibits a significant amount of
variability, comparable to that of either of its component members.

\begin{figure*}
\centering
\includegraphics[width=0.485\textwidth]{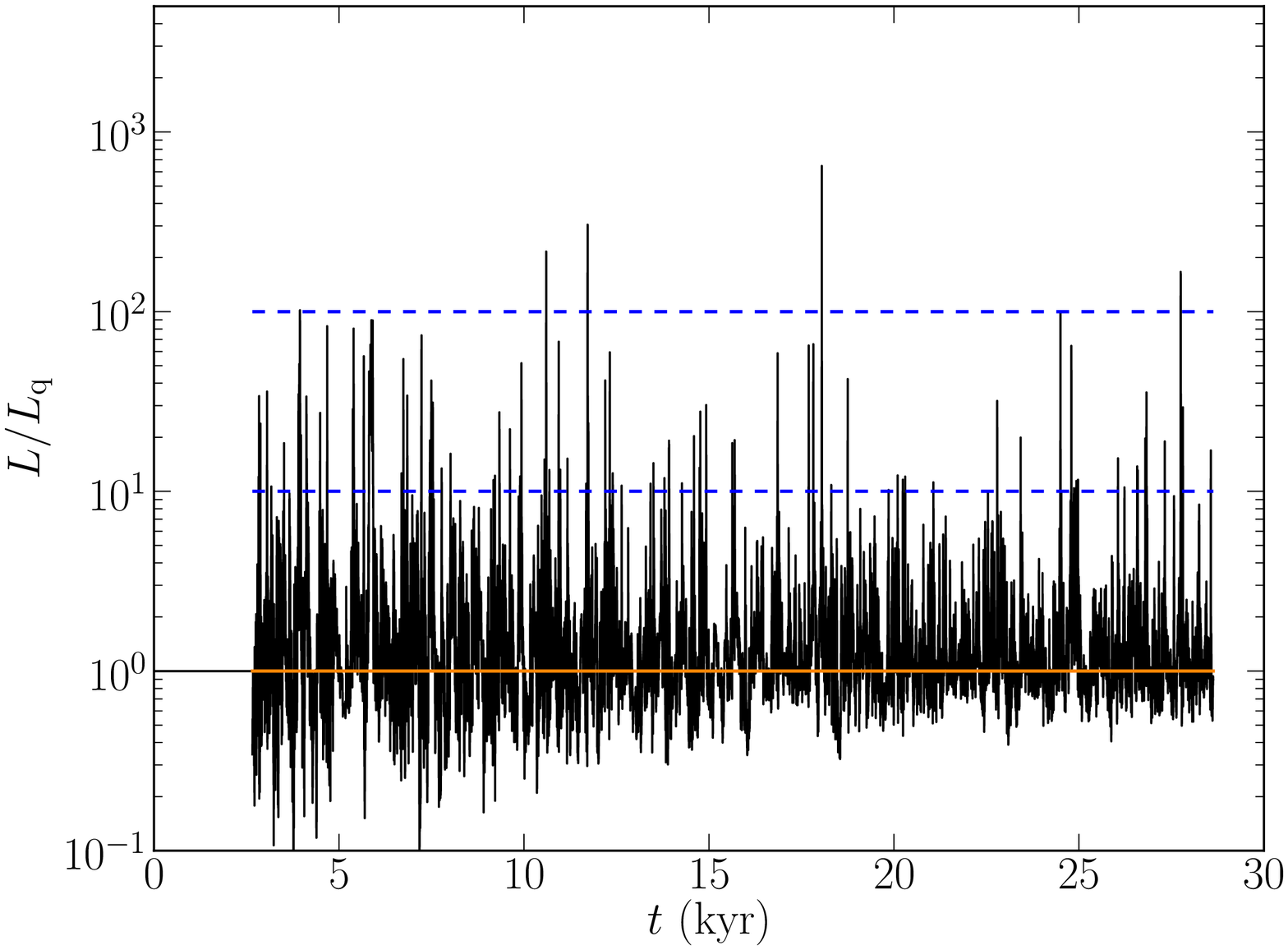}
\includegraphics[width=0.485\textwidth]{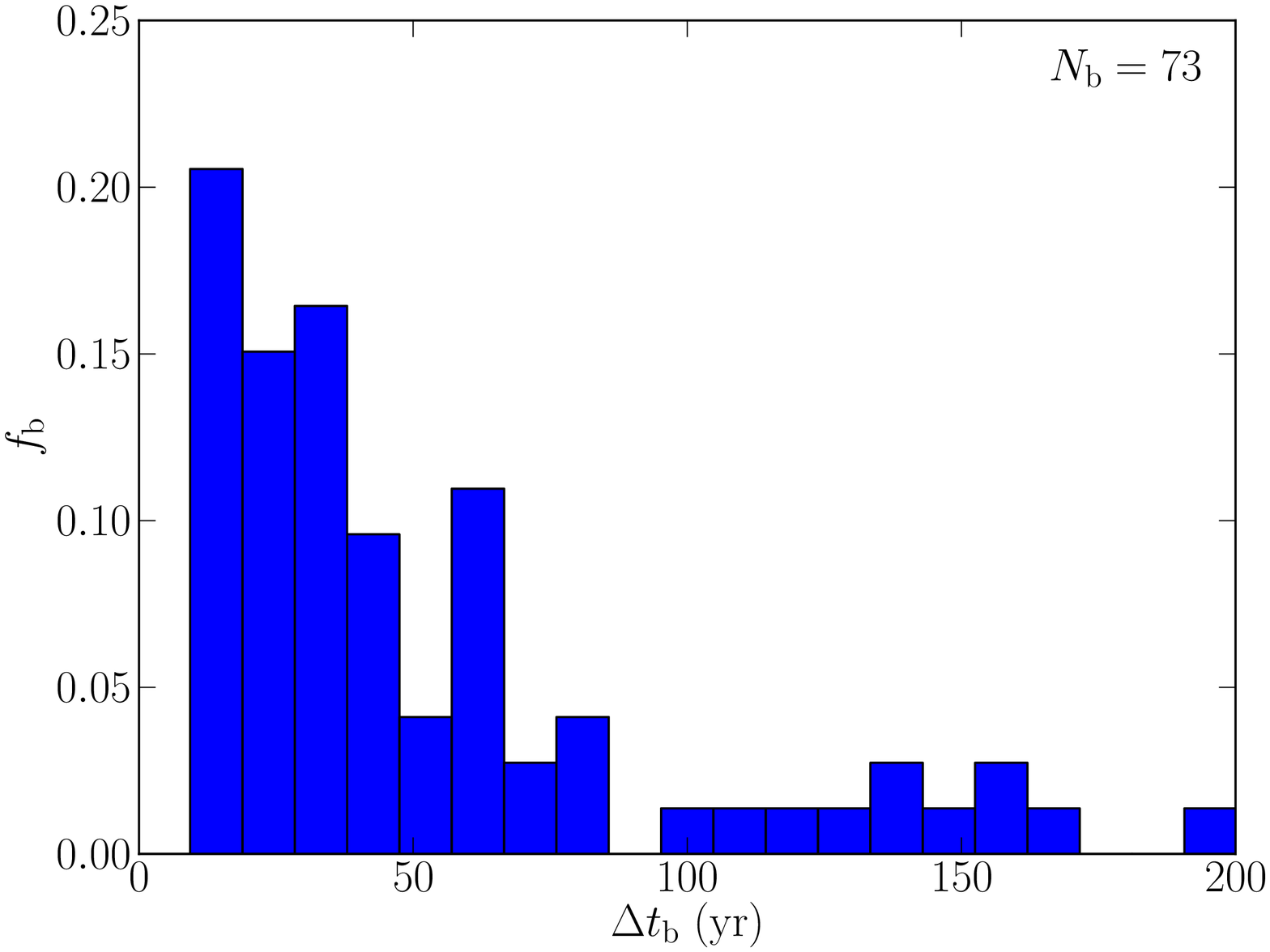} \\
\includegraphics[width=0.485\textwidth]{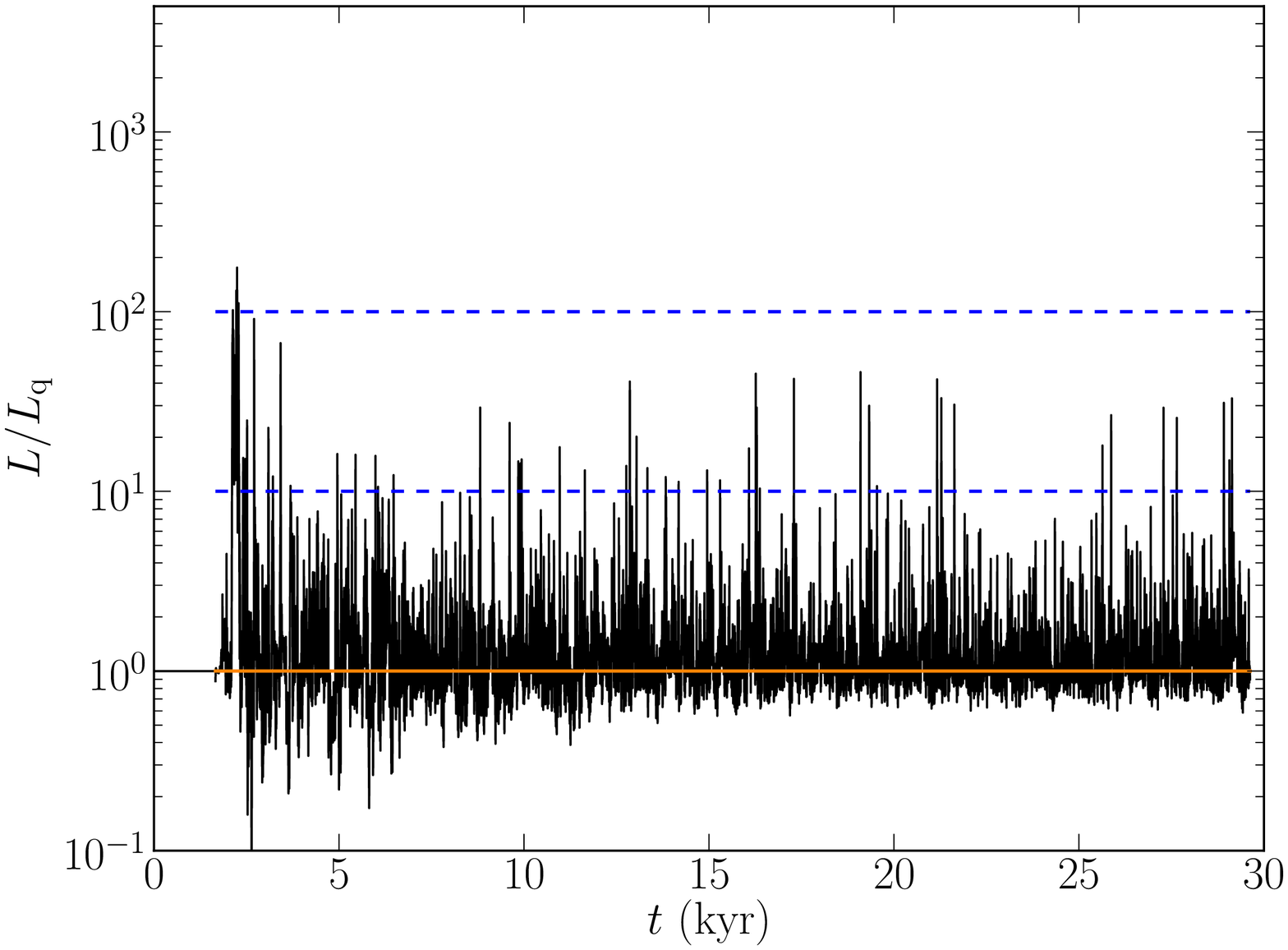}
\includegraphics[width=0.485\textwidth]{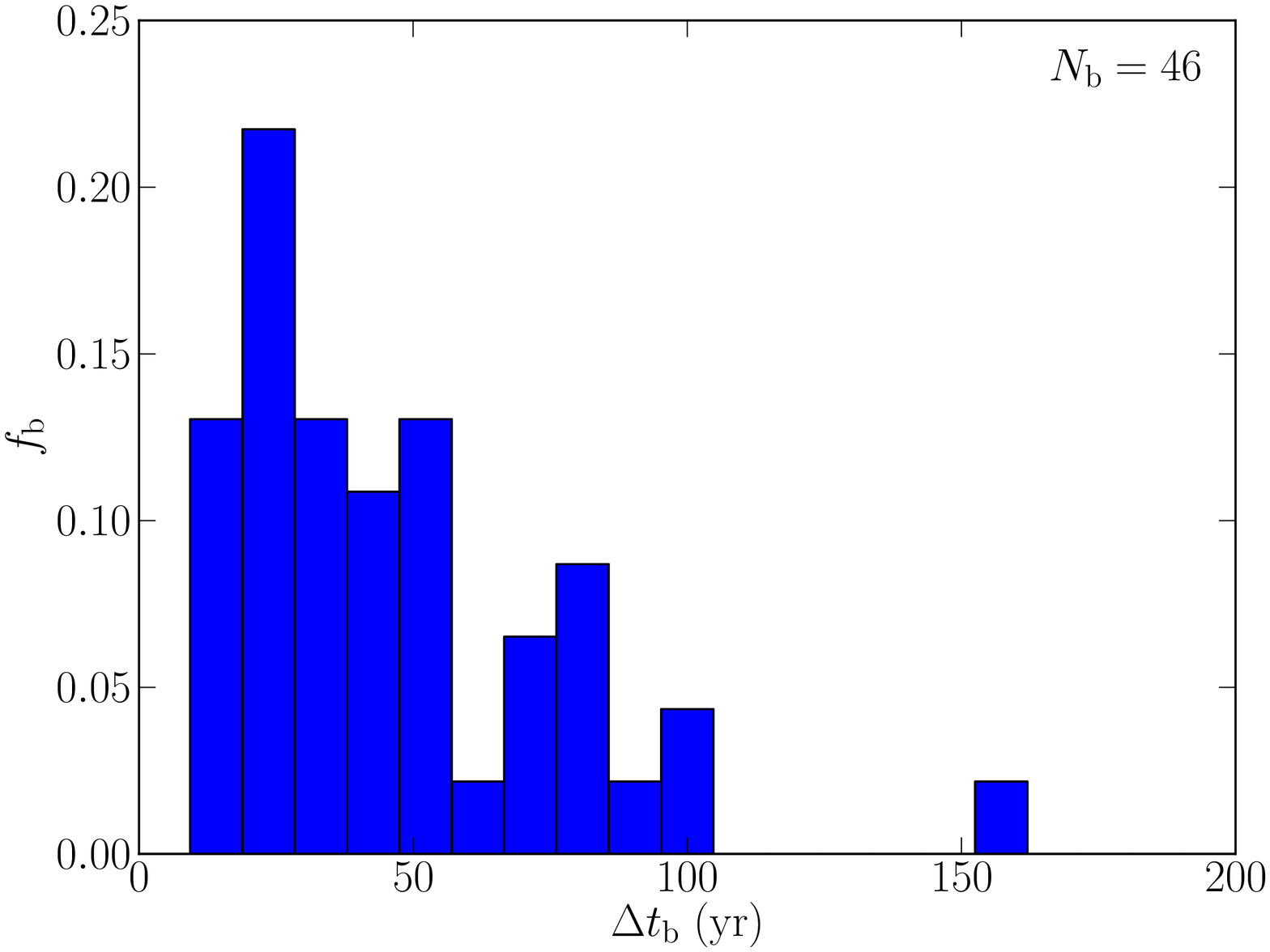}
\caption{Normalized accretion luminosities $L/L_{\rm q}$ (left column) and
  histograms of the fractional number distributions of burst durations
  (right column) for two additional clusters containing $N = 16$
  (top), and $N = 128$ (bottom) members. The dashed blue lines in the left
  panels denote thresholds of $10$ and $100{\times}$ the quiescent
  luminosity; all other lines and colors are as they appear in Figure
  \ref{fig:N=2clusterexample}. The normalized quiescent luminosity in
  each case corresponds to approximately $4 \times 10^5$ and $4 \times
  10^6\,\LSun$, respectively.  During the evolution of the $N = 16$
  cluster a confluence of burst events at $t \approx 18\,\mbox{kyr}$
  produce a particularly prominent luminous event that is
  $647{\times}$ more luminous than the cluster's quiescent
  level. However, the competition between the increasing cluster
  quiescent luminosity $L_{\rm q}$ and the potential for such
  overlapping bursts is evident in the reduced magnitude of the
  fluctuations above $L_{\rm q}$ with increasing cluster size $N$. 
  In the right panels, the total number of identified bursts ($N_{\rm b}$) 
  is indicated in the upper-right.}
\label{fig:normLcomparison}
\end{figure*}

In the right-hand column of Figure \ref{fig:N=2clusterexample} we
present histograms of the duration of the burst events exhibited by
the individual cluster members (top- and middle-right), and for the
cluster as a whole (bottom-right). The first cluster member exhibits a
total of $73$ individual burst events over the course of
${\sim}20.5\,\mbox{kyr}$---the span of time from when the disk forms
to when the protostar reaches a mass of ${\sim}40\,\MSun$ and the
simulation is terminated. The mean burst duration is found to be $\sim
96.3\,\mbox{yr}$. This amounts to approximately $7.0\,\mbox{kyr}$ that
are spent exclusively in the burst mode of accretion, with the
remaining $13.5\,\mbox{kyr}$ spent in quiescent phases. By comparison,
during the approximately $18.5\,\mbox{kyr}$ from the time the second
cluster member forms to when the simulations end, the second cluster
member exhibits a total of $68$ individual burst events. The typical
burst duration is found to be ${\sim}80.2\,\mbox{yr}$, for a total of
${\sim}5.5\,\mbox{kyr}$ spent accreting via the burst mode, and the
remaining roughly $13.0\,\mbox{kyr}$ spent in quiescent phases.

From this we can conclude that the burst mode of accretion plays a
significant role in the mass growth of Population III protostars. In
both cases about $30\%$ of each protostar's individual accretion
history is spent in the burst mode---that is to say that
gravitational-instability--driven fragmentary accretion is responsible
for one-third to one-half of the total accretion onto the first
stars. This is actually a significantly higher proportion of time than
has been observed in analogous simulations of present-day star
formation, wherein the burst mode of accretion is thought to be
responsible for only ${\sim}10\%$ of a protostar's accretion history
(e.g., \citealt{vorobyov2006,vorobyov2010};
\citetalias{vorobyov2013}).

The number and frequency of the observed bursts is also affected by
whether one considers each cluster member individually or simply
considers the cumulative luminosity of the cluster as a whole. The
first and second cluster member yield $73$ and $68$ burst events over
$18.5$ and $13.0\,\mbox{kyr}$, respectively---$1$ event roughly every
$200\,\mbox{yr}$ in both cases. Performing the same analysis of the
luminosity, but of the cluster as a whole, the perceived number of
burst events would be only $74$: far fewer than might be expected
given the number of actual burst events experienced by the cluster's
component members.

Hence, the quiescent mode is clearly the dominant mode of
accretion. As it is unlikely for all members of a cluster to be
dimming simultaneously, the variations below the quiescent rate in the
luminosity of a single cluster member rate are largely obscured. The
cluster luminosity is thus somewhat greater than might be suggested by
a simple summation of the individual luminosities. Hence, the number
of burst events as determined by examining the cluster luminosity is
actually fewer than might be expected, because individual burst events
can only be identified when the luminosity of an individual star
exceeds the luminosity of the cluster as a whole. As a result, the
perceived duration of burst events also decreases.

\begin{figure}
\centering
\includegraphics[width=\columnwidth]{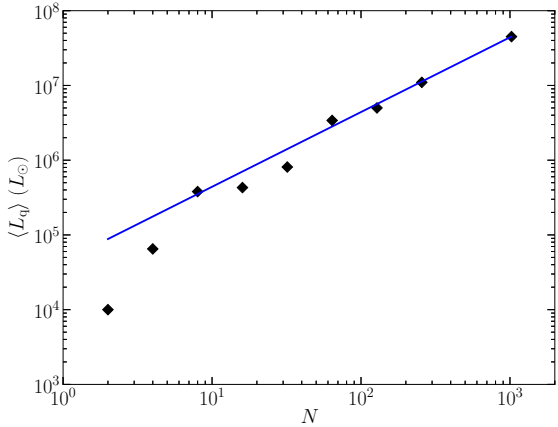}
\caption{The time-averaged quiescent luminosity $\langle L_{\rm q}
  \rangle$ in clusters of size $N \in \{ 2, 4, 8, 16, 32, 64, 128,
  256, 1024 \}$ (from left to right) represented by diamond
  symbols. $\langle L_{\rm q} \rangle$ increases roughly linearly as a
  function of the number $N$ of cluster members as shown by the best
  fit line.}
\label{fig:straightline}
\end{figure}

To compare the variations in the luminosities of clusters of different
size $N$, we normalize each cluster luminosity $L$ by its own
quiescent rate $L_{\rm q}$, and in Figure \ref{fig:normLcomparison}
compare them for clusters with $N = 16$ and $N = 128$. Luminosities
are plotted in black; quiescent cluster luminosities in orange; and
the dashed blue lines denote luminosity levels $10\times$ and
$100\times$ the quiescent rate $L_{\rm q}$. The panels show that for
larger $N$ it is less likely that there are burst events for which the
cluster luminosity reaches above a certain multiple of $L_{\rm
  q}$. Conversely, as $N$ increases, there is also the possibility
that two or more cluster members will experience simultaneous
independent burst events. We use Figure \ref{fig:straightline} to
illustrate how the the time-averaged quiescent luminosity $\langle
L_{\rm q} \rangle$ increases with the number $N$ of cluster
members. The best-fit line represents a linear relationship. For
completeness, we model clusters as large as $N=1000$, however most
estimates of Population III clustering would imply much smaller
clusters, as discussed earlier in this paper.

\begin{figure}
\centering
\includegraphics[width=\columnwidth]{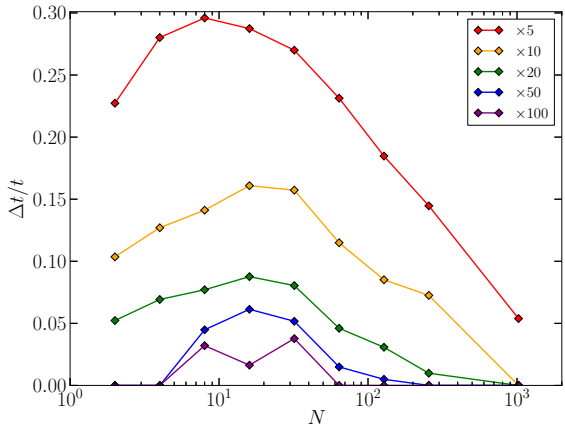}
\caption{The fractional duration of total time, $\Delta{t}/t$, that a
  a cluster of a given size $N$ spends at an elevated
  luminosity. Shown are the values of $\Delta{t}/t$ for different
  values of elevated luminosity $L/L_{\rm q} = 5, 10, 20, 50$, and
  $100$ (appearing sequentially top to bottom). Most curves shows a maximum at $N \simeq 16$. Larger clusters
  exhibit decreasing variability above the mean quiescent rate
  $\langle L_{\rm q} \rangle$.}
\label{fig:clusterbursts_durationvsN}
\end{figure}

In Figure \ref{fig:clusterbursts_durationvsN}, the vertical axis shows
the duration of time ${\Delta}t$ as a fraction of the cluster's total
star-forming lifetime $t$ that a cluster spends above a certain
luminosity threshold. The horizontal axis shows the sizes $N \in \{2,
4, 8, 16, 32, 64, 128, 256, 1024\}$ of the individual primordial
star-forming clusters. Each curve represents a specific luminosity
threshold: from top to bottom, of $5, 10, 20, 50, \mbox{and} 100$
times each specific cluster's quiescent luminosity. The diamond
symbols along a single value of $N$ thus indicate how much time that
cluster spends at an elevated luminosity $L/L_{\rm q}$. As the typical
time frame for an individual burst event is on the order of a few
hundred yr, the likelihood of two or more protostars within a cluster
expressing simultaneous burst events is quite low. As the number of
cluster members increases however, such coincidences become
increasingly likely. However, this effect is counteracted by the fact
that each additional protostar in a cluster also contributes to
increasing the mean cluster luminosity. This prohibits all but the
most intense burst accretion events from being visible. A balance
between these effects is achieved for relatively small clusters, $N
\simeq 16$. For these cases, we find nearly $15\%$ of the cluster's
star forming period is spent at a luminosity level $10{\times}$
greater than the mean rate, $L_{\rm q}$. This corresponds to an
absolute luminosity approaching $10^8\,\LSun$.

\section{Discussion \& Conclusions}
\label{sec:dnc}

We have presented a scenario for the assembly of a cluster of first
stars formed from the pristine gas that is pooling into newly
virialized dark matter halos at redshift $z \sim$ 20--50. The Jeans
criterion provides a context for the combination of self-gravity and
anisotropic structure to give rise to fragmentation of the gas. We
estimate that a typical dark matter halo with mass between $10^5$ and
$10^6\,\MSun$ contains $10\%$ by mass of gas, of which roughly $10\%$
actually goes into stars. The Jeans criterion then implies the
formation of clusters containing tens of members (perhaps 10--50
protostars), though the specific numbers are difficult to estimate due
to the ambiguity with which the dispersal and merger of clumps prior
to their collapse occurs \citep[e.g.,][]{greif2011}.

We employ numerical hydrodynamics simulations in the thin-disk limit
to then investigate the long-term evolution of the individual
clumps. We model their collapse self-consistently into the formation
of the protostar and its surrounding disk. In each case a disk forms
relatively quickly, regardless of the initial conditions, within a few
kyr of the formation of the protostar. The disk is centrifugally
supported, but its mass lends it to being gravitational unstable. We 
observe a rapid and episodic fragmentation of the disk during which
fragments having between 0.1--1.0 $\MSun$ are formed at typical
radii of ${\sim}50\,\mbox{AU}$. The fragments are torqued inward
toward the central protostar where they are then accreted, resulting
in massive bursts of luminosity that can exceed the mean rate by as
much as two orders of magnitude, occasionally exceeding $10^6\,\LSun$
(as in Figure \ref{fig:N=2clusterexample}).

In the context of this formation scenario we analyze the luminosity
profiles that are produced by a multiplicity of first stars having
formed approximately coevally within a young cluster. We assume that
each cluster member forms independently and harbors its own disk that
is subject to gravitationally induced episodic fragmentation and
accretion (i.e., the burst mode of accretion). With increasing numbers 
of cluster members, the quiescent luminosity of the cluster is seen to
steadily increase. However, increases in the number of cluster members
also increase the probability that two or more members experience a
burst of accretion simultaneously. In one cluster simulation we
observe one such particularly luminous super-posed accretion event
during which the cluster luminosity rises to nearly $1000{\times}$
it's quiescent rate, to $2.48{\times}10^8\,\LSun$ (Figure
\ref{fig:normLcomparison}).

Competition between these two effects results in clusters of a certain
size ($N \simeq 16$) spending a sizable fraction of their star
forming life-time (roughly $15\%$) at luminosities $10{\times}$
greater than might otherwise be expected with smooth accretion
\citep[e.g.,][]{smith2011}. Each additional object added to a cluster
raises the mean/quiescent luminosity of the cluster, and this in turn
tends to suppress the prominence of individual and multiple burst
events in clusters in which $N \gtrsim 16$. Such moderately sized
clusters are also the most easily manufactured, capable of being
produced assuming even a relatively low star formation efficiency, and
are of a number as has been counted in a variety of theoretical
Population III star forming scenarios to date
\citep[e.g.,][]{stacy2010,clark2011b,greif2011}.

The possible formation of clusters of Population III stars in the
dark matter halos of the early universe provides a unique opportunity
for observations. \citet{rydberg2013} estimated the total luminosity
from even the most upper mass estimates of Population III stars (i.e.,
those with masses ${\sim}300\,\MSun$) and found that even these were
likely too faint to be observed by next-generation telescopes such as
the James Webb Space Telescope. The likelihood for such instruments to
observe even lower mass Population III stars is thus highly
improbable. However, as we have shown, clusters of even low mass
Population III stars (as others have also indicated may exist: Bromm
et al.~2003; Clark et al.~2008; and Greif et al.~2012), are capable of
producing luminosities in excess of lone very massive Population III
stars (for example \citet{bromm2001} estimate luminosities in the
range $\simeq 10^6$--$10^7\,\LSun$ for masses in the range 100--500
$\MSun$). Observational constraints on the early stages of 
reionization, from $21\,\mbox{cm}$ observations by the Square
Kilometer Array (e.g., Carilli et al.~2004) may actually provide
indirect constraints on the abundance of such clusters. Observations
by the Atacama Large Millimeter/submillimeter Array (ALMA) of emission
from dusty galaxies in the early universe may also provide insight
into their possible formation as influenced by the formation of an
initial cluster of Population III stars. However, it will be in the
next decade that next-generation telescopes may actually provide
direct constraints on the Population III star formation taking place
in the early universe; with clusters of Population III stars being
some of the most luminous at that epoch.


\section*{Acknowledgements}
\label{sec:acknowledgements}

We thank Eduard Vorobyov for early discussions and the anonymous referee
for comments that improved the presentation of the paper.
SB acknowledges support from a Discovery Grant from the Natural
Sciences and Engineering Research Council (NSERC) of Canada. This
research has made use of NASA's Astrophysics Data System.

\bibliography{myrefs}



\end{document}